%% file: 2body.tex
\documentstyle[epsf,nomargid,a4wide,12pt]{article}
\input{laa}

\newcommand{\Msol}{\mbox{M$_{\sun}$}}

\newcommand{\eg}{{e.g.,} \thinspace}
\newcommand{\ie}{{i.e.,} \thinspace}

\title{Two--body heating in numerical\\ galaxy formation experiments}
\author{Matthias Steinmetz$^{1,2}$ and Simon D.M.~White$^1$\\
$^1${\small Max--Planck--Institut f\"ur Astrophysik, Postfach 1523, 85740 Garching, Germany}\hfill\\
$^2${\small Department of Astronomy, University of California, Berkeley, CA 94720, USA}\hfill\\}
\date{\small Accepted . Received ; in
original form 1996 September 3}
\begin{document}
\maketitle
\begin{abstract}
We show that discreteness effects related to classical
two-body relaxation produce spurious heating of the gaseous component
in numerical simulations of galaxy formation. A simple analytic calculation
demonstrates that this artificial heating will dominate radiative
cooling in any simulation where the mass of an individual dark matter
particle exceeds a certain critical value. This maximum mass depends only
on the cooling function of the gas, on the fraction of the material
in gaseous form, and (weakly) on typical temperatures in the gas. It
is comparable to, or smaller than, the dark matter
particle masses employed in most published simulations of
cosmological hydrodynamics and galaxy
formation. Any simulation which violates this constraint
will be unable to follow cooling flows,
although catastrophic cooling of gas may still occur in regions with
very short cooling times. We use a series of N--body/smoothed particle
hydrodynamics simulations to explore this effect. In simulations which
neglect radiative cooling, two--body heating causes a gradual
expansion of the gas component. When radiative effects are included,
we find that gas cooling is almost completely suppressed for dark
matter particle masses above our limit.
Although our test simulations use smoothed particle hydrodynamics,
similar effects, and a similar critical mass, are expected in any
simulation where the dark matter is represented by discrete particles.
\end{abstract}
\section{Introduction}
Over the last twenty years our understanding of structure formation
has benefitted substantially from numerical N-body simulations. It is
now clear that such simulations provide a robust and efficient, though
sometimes computationally expensive tool to obtain approximate
solutions for the evolution of a collisionless self-gravitating
``fluid'' from cosmologically relevant initial conditions
(Efstathiou et al 1985).  Although such cosmological N--body 
simulations can now be performed using particle numbers in excess of
$10^7$, and can follow density contrasts over a range of $10^6$, this
still turns out to be barely sufficient to address questions related
to the formation of galaxies in a proper cosmological context. Current
observational data appear to favour models in which structure is built up
hierarchically, and so it is of particular interest to ask how
numerical resolution affects the mass hierarchy at the low mass end.
It appears that while the detailed inner structure of halos can only
be analyzed reliably if they contain at least several hundred particles,
the distribution of halos and their total masses are reasonably
represented for halos with ten or more particles (e.g. Efstathiou et
al 1988). This convergence is of major importance because it means
that simulations performed with $\sim10^7$ particles can reliably
resolve the halos of galaxies like the Milky Way while simultaneously
covering a large enough region (a few hundreds of Mpc) to allow a
statistical comparison with galaxy surveys. The most serious
limitation of such models is the fact that the formation of real galaxies
was clearly strongly affected by a wide variety of physical processes
which are not included in the simulations.

Within the last few years it has become possible to include a number
of additional 
processes in such models by following, in addition to the dark matter,
a dissipative gaseous component. However, even in the simplest case of a 
nonradiative, non--star--forming gas, the combined dark matter/gas
system exhibits much more complex behaviour than a pure N--body
system. In addition, such simulations are more CPU--intensive and so
typically have poorer resolution than the best N--body simulations. 
Our current understanding of the effects of this limited resolution is
still quite rudimentary. Systematic analyses of the convergence of such 
simulations are only just
beginning (Kang et al 1994; Frenk et al, in preperation).
In this paper we investigate how the finite dark
matter particle mass affects the dynamics and the cooling capabilities of the
gas. We show that discreteness effects
give rise to a steady energy flow from the dark matter to
the gas. This heating is strong enough to affect the structure of the
gas within any halo made up of fewer than 1000 dark matter particles.

The outline of our paper is the
following. In the next section we derive an analytic formula for the
rate at which discreteness effects transfer energy from the dark
matter to the gas. We compare this with the expected radiative cooling
rate, we discuss how these rates scale with physical and simulation
parameters, and we draw some first conclusions about how numerical 
simulations should be designed. In section 3 we check this analytic 
theory using a set of numerical simulations based on smoothed particle
hydrodynamics. Simulations with and without radiative cooling are 
investigated separately. Section 4 discusses and summarizes our results.
\section{Analytic theory}
Consider a fluid element of mass $m_{\rm g}$ and density $\varrho_{\rm
g}$ which is at rest. This fluid element encounters a dark matter particle
of mass $M_{\rm DM}$ and relative velocity $v$ with a closest approach
distance $b$. In the impulse approximation (see, \eg Binney \&
Tremaine 1987), the fluid element is accelerated to velocity
\begin{equation}
\Delta v = \frac{2\,G\,M_{\rm DM}}{b\,v}\, ,
\end{equation}
or to a corresponding kinetic energy
\begin{equation}
\Delta E = \frac{2\,G^2\,M^2_{\rm DM}\,m_{\rm g}}{b^2\,v^2}\, .
\end{equation}
This energy is dissipated to heat by shocks, by artificial
viscosity, or by an adiabatic expansion of the
gas to a new equilibrium state. Such encounters
occur with a rate $2\pi\,v\,b\,db\,n_{\rm DM}$, so the heating rate
can be written as
\begin{equation}
\left.\frac{dE}{dt}\right|_{\rm heat} =
\int\,d^3v\,f(v)\,\int_{b_{\rm min}}^{b_{\rm max}}\,2\,\pi\,db\,
\frac{2\,G^2\,M_{\rm
DM}\,\varrho_{\rm DM} m_{\rm g}}{b\,v}\, ,
\end{equation}
where $f(v)$ is the velocity distribution function for the dark matter
particles. Assuming this to be Maxwellian, we obtain after
the evaluation of the integrals
\begin{equation}
\left.\frac{dE}{dt}\right|_{\rm heat} =
\sqrt{\frac{32\,\pi}{3}}\,G^2\,\ln\Lambda\,\frac{M_{\rm DM}\,m_{\rm
gas}\,\varrho_{\rm DM}}{\sigma_{\rm 1D}}\, ,
\end{equation}
$\sigma_{\rm 1D}$ being the 1D velocity dispersion of the dark matter 
and $\ln\Lambda$ the Coulomb logarithm. For typical galaxy formation 
experiments $\ln\Lambda$ is in the range 3 to 7. 

In an equilibrium system the internal energy of the fluid element is
$E\sim 3 m_{\rm g}\sigma_{\rm 1D}^2 /2$, so we can define a
characteristic heating time by $t_h = E\big/ (dE/dt)$, or
\begin{equation}
t_h =
\sqrt{\frac{27}{128\,\pi}}\,\frac{\sigma_{\rm
1D}^3}{G^2\,\ln\Lambda\,M_{\rm DM}\,\varrho_{\rm DM}}\, .
\end{equation}
For an equilibrium dark matter dominated halo we can get a more
transparent formula by comparing the heating rate near the halo
half-mass radius $R_h$ to the circular orbit period at this radius 
$t_c$. For a halo made up of $N$ dark matter particles in total, the 
definitions and approximate relations, $N\,M_{\rm DM}/2\approx 2 
R_h\,{\sigma_{\rm 1D}^2/G}$, $\varrho_{\rm DM}\approx N\,M_{\rm DM}/8\,\pi\,R_h^3$,
and $t_c = 2\,\pi\,R_h/\sqrt{2}\sigma_{\rm 1D}$, allow us to cast equation
(5) in the form
\begin{equation}
\frac{t_h}{t_c} =
\sqrt{\frac{3}{\pi}}\,\frac{3\,N}{32\,\ln\Lambda}.
\end{equation}
Thus for halos with around 50 dark matter particles the heating time
for gas at the half-mass radius is comparable to the orbital
period at that radius, and so to the halo formation time. The 
gas distribution within such halos will clearly never be free from
substantial two-body heating effects. In a cosmological
simulation, even a halo of $10^3$ dark matter particles will typically
have been around for several nominal formation times, 
and so will suffer 10 to 20\% effects near its half-mass radius.
Because of the strong $\varrho_{\rm DM}$ dependence of equation (5),
effects in the inner regions will be substantially stronger.
A worrying aspect of these results is that in hierarchical
clustering all massive objects build up through the
aggregation of smaller collapsed systems; two-body heating must be
important in the first generations of halos which form in any
simulation, and it is unclear how well such early artifacts will be
eliminated by later evolution.

Let us now compare the two-body heating rate with the radiative
cooling rates expected for the gas in a realistic simulation:
\begin{equation}
\left.\frac{dE}{dt}\right|_{\rm cool} = \varrho_{\rm g}\,m_{\rm g} (N_{\rm
A}\,X_{\rm H})^2 \Lambda(T)
\end{equation}
$N_{\rm A}$, $X_{\rm H}$, and $\Lambda(T)$ being Avogadro's constant,
the total mass fraction of hydrogen and the cooling function,
respectively. Comparing  equations (4) and (7), two-body heating will 
dominate over radiative cooling, if
\begin{equation}
\sqrt{\frac{32\,\pi}{3}}\frac{G^2\,\ln\Lambda\,M_{\rm DM}\,
\varrho_{\rm DM}}{\varrho_{\rm g}(N_{\rm A}\,X_{\rm H})^2
\Lambda(T)\sigma_{1D}}
> 1\, .
\end{equation}
Since heating and cooling are two--body processes they both scale 
as $\varrho^2$. If we define $f\equiv\varrho_{\rm g}/\varrho_{\rm DM}$ 
and use convenient units for other quantities, we find that this
inequality is equivalent to
\begin{equation}
M_{\rm DM} < M_{\rm crit} \equiv
4~10^9\,\sigma_{100}\,(\ln\Lambda)_5\,f^{-1}_{0.05}\,\Lambda_{-23},
\end{equation}
where $\sigma_{100}$ is the 1D velocity dispersion in units of
100\,km\,s$^{-1}$,
$f_{0.05}$ the local baryon fraction divided by 0.05, $(\ln\Lambda)_5$ the
Coulomb logarithm divided by 5, and $\Lambda_{-23}$ the cooling function in
units of $10^{-23}\,$erg\,cm$^{-3}$ per (H atom cm$^{-3}$)$^2$,
respectively; here we have assumed $X_{\rm H} = 0.76$.
Identifying $\sigma_{1D}$ with the
corresponding virial temperature, \ie $\sigma_{100}=1.2\sqrt{T_6}$, 
where $T_6$ is temperature in units of $10^6$\,K, we can write
this critical mass in the alternative form
\begin{equation}
\label{masscrit}
M_{\rm crit} =
5~10^9\,\Msol\,\sqrt{T_6}\,(\ln\Lambda)_5\,f^{-1}_{0.05}\,\Lambda_{-23},
\end{equation}
At first sight it is surprising that our critical mass turns out to be 
of galactic scale; only atomic and gravitational constants contribute
to the right-hand-side of the inequality in equation (8), and we will
see that the temperature dependence of equation (10) is quite weak
over the range of interest.  This ``coincidence'' reflects the 
well-known fact that the characteristic masses of galaxies
appear to be determined by the condition that the cooling time for gas
in a protogalactic dark halo should be comparable to the halo
formation time (\eg Rees \& Ostriker 1977; White \& Rees 1978).
It is important to note that equation (9) is purely local and makes
no specific assumptions about hydrostatic equilibrium or about the
relative distributions of gas and dark matter. Only in equation (10)
do we implicitly adopt such assumptions when we identify the gas
temperature with the dark matter velocity dispersion. In practice, 
this is a weak assumption which is approximately correct in most
situations of interest.  We also note that our derivation is not 
specific to any particular numerical treatment of hydrodynamics; 
it depends only on the assumption that the dark matter is represented 
by particles of mass $M_{\rm DM}$. Our critical mass should be
relevant for almost all the numerical methods currently in use to
carry out cosmological hydrodynamics simulations.

The arguments given above apply only if the local cooling time is 
comparable to or longer than the local dynamical time. If cooling 
can occur on a much shorter timescale, the gas will lose its
internal energy faster than the typical encounter time and two-body
effects will be unable to reheat it. Such ``catastrophic'' cooling is 
unaffected by the process we are discussing. Another complication, 
which we will not discuss further, is that the energy deposited by an 
encounter may not produce local heating, but may be transported by
sound waves to other regions of the system before it is dissipated.

On the basis of this analysis, we can make the following predictions for
galaxy formation experiments:
\begin{itemize}
\item In simulations where radiative cooling is not included, energy 
will be steadily transferred from the dark matter to the gas. This
will lead to a gradual expansion of the gas component in supposedly
equilibrium systems.
\item In simulations which include radiative processes but
where the mass of a dark matter particle exceeds the critical value
of equations (9) and (10), cooling will be suppressed 
wherever the cooling time is comparable to
or longer than the local dynamical time (\ie in the cooling flow regime).
\end{itemize}

\begin{figure}
\mbox{\epsfxsize=1\hsize\epsffile{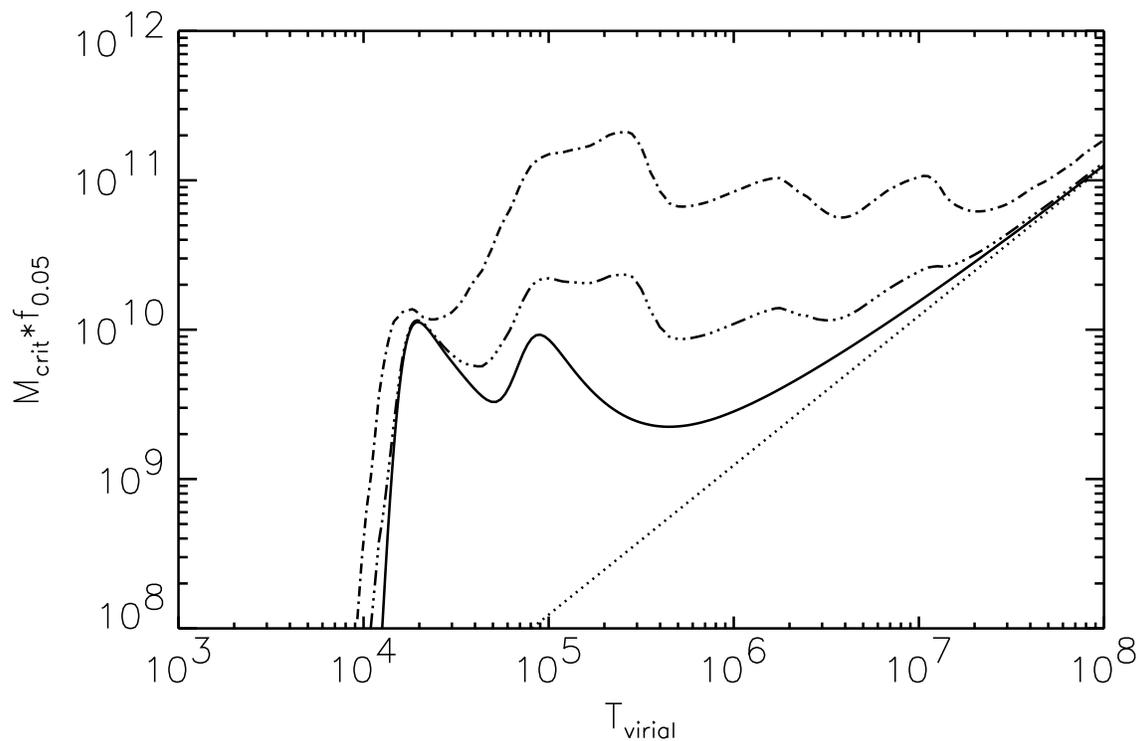}}
\caption[]{\label{colfig}Critical mass for dark matter particles
(according to equation (\ref{masscrit})) as
a function of the virial temperature of the simulated system.  The 
different curves correspond to different cooling functions: 
primordial composition (solid line), solar metallicity (dashed--dotted),
$\frac{1}{10}$ of solar metallicity (dashed--triple dotted), and pure
bremsstrahlung (dotted).}
\end{figure}

Let us identify $T$ with the virial temperature of a dark matter halo
and assume gas and dark matter to be distributed similarly.
For a given cooling function, we can then use equation (10) to calculate 
the critical mass of dark matter particles. Figure 1 shows the result 
as a function of $T$. These calculations assume a baryon fraction of
5\% but are easily scaled to other values. For a cooling function 
appropriate to solar metallicity gas, the critical
dark--matter particle mass is $\sim 10^{11}\,$M$_\odot$ for all temperatures
between $10^5$ and $10^8\,$K, hence for objects ranging from dwarf
galaxy halos to rich galaxy clusters. Similarly, for a metallicity
of $0.1\,Z_\odot$, the mass critical mass is $\sim 10^{10}\,$M$_\odot$ for
$T$ between $2\times 10^4$ to $5\times 10^6\,$K, the whole range
relevant to galaxy halos. We therefore come to the surprising
conclusion that one should use the same dark matter particle mass 
when simulating small galaxies as when simulating
galaxy clusters. For gas of primordial composition, a realistic
galaxy/cluster formation simulation requires two--body heating to
be unimportant for any object with a virial temperature in the range
$10^5$ to $10^8$\,K, implying that the dark matter particle mass should not
exceed about $2~10^9$\,M$_\odot$. A simulation of a rich cluster would
then need about half a million particles within the virial radius, 
a criterion which is failed by all simulations published so
far. Nevertheless, for cluster simulations with several thousand 
particles, two--body heating times are comparable to the Hubble time 
only in the inner regions, so it is possible that only the core
structure of the gas is affected by numerical artifacts.
Figure 1 also shows a cooling function due to bremsstrahlung alone. 
This approximates the extreme case of cooling in the presence of a
strong UV background, where collisionally excited line emission can 
be almost completely suppressed (\eg Efstathiou 1992). In this case, 
$\Lambda\propto \sqrt{T}$, and so
$M_{\rm crit}\propto T$. For a  dwarf galaxy ($T_{\rm vir} = 10^5\,$K),
dark matter particle masses are required to be below $10^8\,\Msol$,
\ie the galaxy halo should be represented by several hundred particles. 

Finally we note that $M_{\rm crit}$ is the value of the dark matter
particle mass for which radiative cooling and artificial two--body
heating are equal. To get realistic results any numerical simulation
should use particle masses which are at least a factor of two or three
smaller than $M_{\rm crit}$.
\section{Numerical verification}
In the previous section we concluded that for parameters
typical of current galaxy formation experiments, two--body heating 
can substantially affect the properties of the gas. We predict that
in simulations without cooling the gas in equilibrium systems will
slowly expand, while in simulations that include radiative losses the
gas may still be prevented from cooling where it should.
In this section we will test these predictions by
means of some numerical experiments.  These were performed using the
smoothed particle hydrodynamics code GRAPESPH (Steinmetz 1996). 
We choose initial conditions which are relevant for galaxy
formation simulations but which also allow easy control over
experimental parameters. In particular, we choose the initial
conditions proposed by Katz (1991): a homogeneous, overdense, and 
rigidly rotating sphere of total mass $8.1\times 10^{11}\,$M$_\odot$ with
superposed small-scale density fluctuations drawn from a CDM
power spectrum. The evolution was simulated with different particle
numbers, $N= 250$, 500, 1000, 4000 and 17000. The initial conditions 
for the lower resolution simulations were obtained by random sampling
those of the model with 17000 particles. This object collapses at
$z\sim 2$ but the simulations were followed over a Hubble time and
the structure was analysed only after $z= 1$ when gas and dark
matter have relaxed and are approximately in equilibrium. 
For simulations including cooling, the cooling was switched on
only after $z = 1$. (Throughout we assume $\Omega=1$ and $H_0=50$
km~s$^{-1}\,$Mpc$^{-1}$.) The density profiles of the relaxed system have
$\varrho\propto r^{-2}$ between 20 and 90\,kpc; the profile is steeper
at larger and shallower at smaller radii, resembling the ``universal''
halo structure proposed by Navarro, Frenk \& White (1996). The gas
temperature is almost uniform within $50\,$kpc, but drops at larger radii.
\subsection{Simulations without cooling}
\begin{figure}
\mbox{\epsfxsize=1\hsize\epsffile{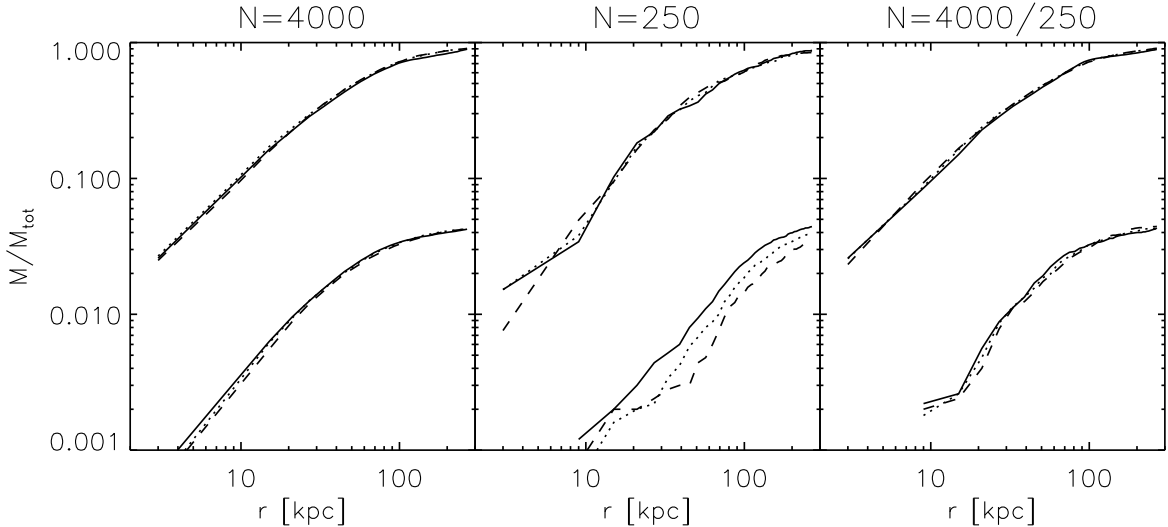}}
\caption[]{\label{profile}Time evolution of the cumulative mass profiles for
dark matter
(upper curves) and gas (lower curves) for three different resolutions: 4000
particles in each component (left), 250 particles in each component 
(middle), and 250 gas particles but 4000 dark matter particles
(right). The curves represent epochs 0~Gyr (solid), 4~Gyr (dotted) and
8~Gyr (dashed) after hydrostatic equilibrium has been established.}
\end{figure}
In Figure \ref{profile} we show the cumulative mass profiles for gas and dark
matter in simulations with differing resolution. The dark matter
profiles show no significant evolution in any of these models. (The
fluctuations at small radii for $N=250$ are just 
statistical noise.) In contrast, the gas distribution expands
substantially for $N=250$, but only slightly for $N=4000$.
To demonstrate that this expansion reflects the mass of the dark
matter particles we also show a simulation with 250 gas
particles and 4000 dark matter particles, \ie gas and dark matter particles
have roughly the same mass. In this case the expansion of the gas 
component is again small, and with the exception of noise effects
at very small radii, the gas profile is
compatible with that obtained using 4000 gas particles.

\begin{figure}[t]
\mbox{\epsfxsize=1\hsize\epsffile{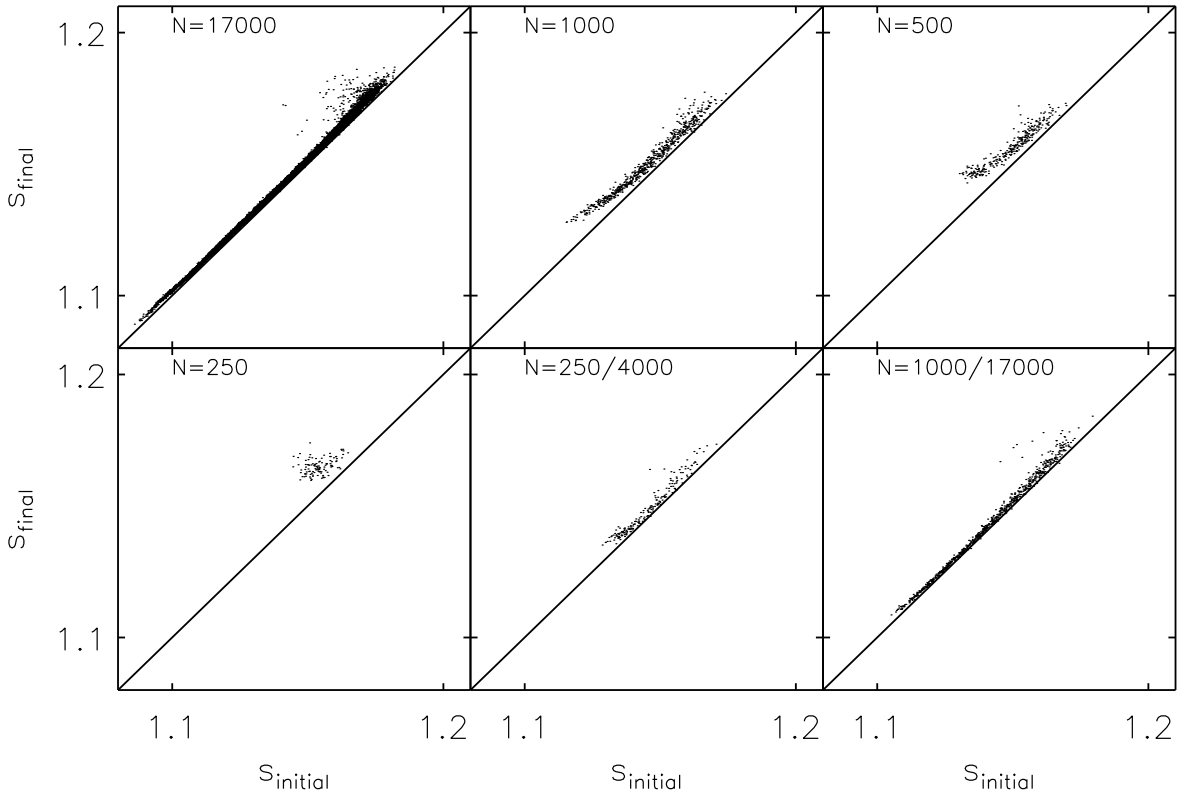}}
\caption[]{\label{entro}Initial versus final specific entropy (in units of
($N_{\rm A}\,k_{\rm b}$)) of gas particles for simulations with
(from top left to bottom right) 17000, 1000, 500, 250 particles in
each of the two components, and with 250/4000 and 1000/17000 
gas/dark matter particles.}
\end{figure}
It is also instructive to look at the heating of the gas in terms of its
entropy
evolution. After hydrostatic equilibrium is established, the specific entropy
$s=\ln(T^{1.5}/\varrho)$ of a gas particle should be constant $\frac{d}{dt}s =
0$. As shown in Figure
\ref{entro}, this is indeed the case for the $N=17000$ run. There is an
almost perfect linear relation between initial and final entropy. This also
demonstrates that for high particle numbers the amount of spuriously generated
entropy due to the artificial viscosity is negligible. Only at very large
radii a slight scatter can be seen, which can be explained by some dynamical
evolution close to
the virial radius. For smaller particle numbers, an evolution in $s$ becomes
visible: when $N=1000$ only particles with the lowest
entropies are affected, while for $N<500$ the whole system
is affected. The low entropy gas exhibits the strongest evolution
because it lies near the center of the halo and so has the shortest
heating time. Comparison with the 250/4000 and 1000/17000 models
shows that entropy generation is not significantly affected 
by numerical resolution in the gas component, but is determined
primarily by the mass resolution of the dark matter component: lowering
the mass of the dark matter particles at constant gas resolution 
suppresses the artificial heating. Note that two--body heating also
occurs when dark matter particle masses are smaller than those of
fluid elements since, unlike the standard stellar dynamical case, the 
fluid elements are at rest when the system is in equilibrium.

\subsection{Simulations with cooling}
We now consider simulations which include radiative cooling. To simplify
comparison of our simulations with the predictions of our analytic
model, we use a schematic cooling function for which $\Lambda={\rm
const}$ above a lower cutoff of $10^4\,$K. As mentioned above, we
switch cooling on only after the initial relaxation phase. Thus the
phase we analyse begins with a relaxed halo, in which the gas has the 
virial temperature and is in hydrostatic equilibrium within the dark 
matter potential well. We choose $\Lambda$ to be
relatively  small so that we can demonstrate two--body heating effects clearly
using about 1000 particles. Statistical noise then has little effect
on our conclusions. These physical conditions are similar to a 
cooling flow. We took the mass of a dark matter particle to be 
$7.7~10^8\,M_{\odot}$, assumed a gas fraction of 5\%, and performed 
two simulations, one with $\Lambda_{-23}=0.4$, the other with 
$\Lambda_{-23}=0.1$. The half-mass radius of the collapsed system is 
then 60\,kpc, while the gravitational softening is 5\,kpc giving
a Coulomb logarithm of about 2.5. According to equation
\ref{masscrit}, for the appropriate virial temperature, $T_6=1.5$, our
chosen particle mass corresponds to the critical value for a 
cooling coefficient of $\Lambda_{-23}=0.25$, midway between those of
our two simulations.  We expect, therefore, that in
one of these two models the gas will be able to cool while in the
other it will not. The two
simulations were run for the same number of nominal cooling time
scales, \ie the model
with the lower cooling coefficient was run for a correspondingly longer
time. In both cases the final cooling radius should be about
23\,kpc implying that about 16 per cent of the gas should be able to
cool.  For an additional comparison we also consider a second model
with $\Lambda_{-23}=0.1$ which differs from the first only in that
the dark matter is represented by 17000 particles. Thus two--body heating 
effects are suppressed by a factor of 17.

\begin{figure}
\mbox{\epsfxsize=0.9\hsize\epsffile{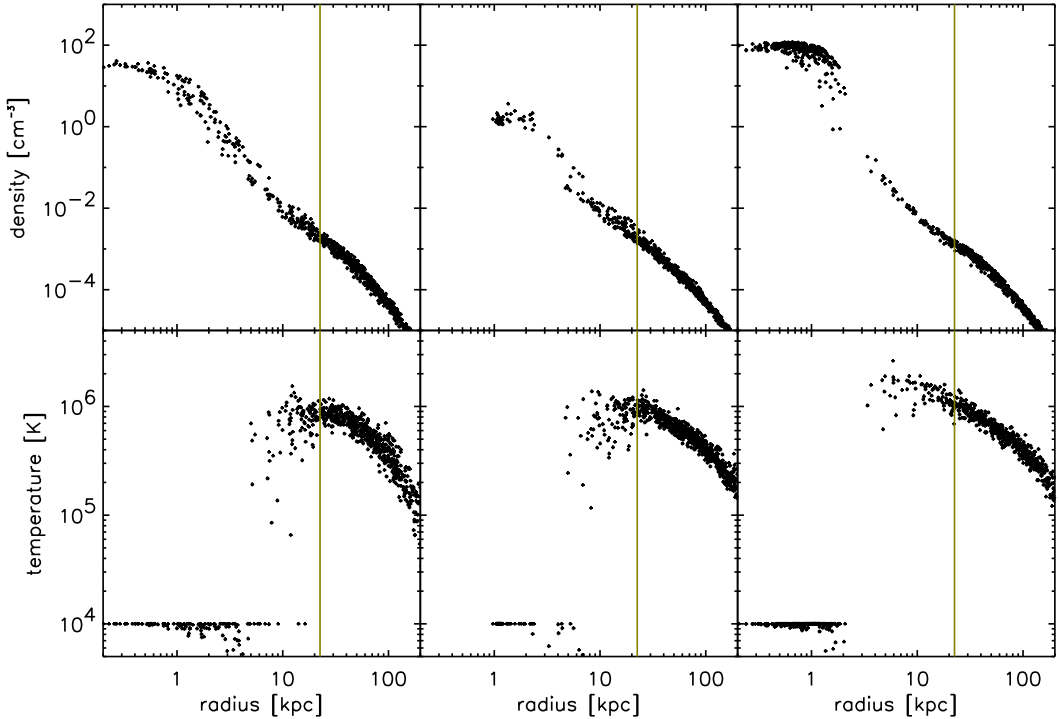}}
\caption[]{\label{colmod}Density (top) and  temperature (bottom) as a function
of radius for two simulations starting from identical initial
conditions but with $\Lambda_{-23}=0.4$ (left) and $\Lambda_{-23}=0.1$ (middle).
These simulations have evolved for the same number of cooling
times since virialisation. The simulation on
the right has the same number of gas particles but 17 times more
dark matter particles. Its evolution time and $\Lambda_{-23}$ value
are identical to those of the model in the middle column. The vertical
dotted line
corresponds to the radius where the cooling time of the initial model 
equals the time for which cooling was allowed.
}
\end{figure}
In Figure \ref{colmod} we show density and temperature profiles for the
final states of the three simulations. In the model with
$\Lambda_{-23}=0.4$ almost all the gas within the cooling radius has started to
cool, and a large fraction has already settled at $10^4$\,K, the
cutoff of the assumed cooling function. The gas distribution responds 
dynamically to this cooling. The density near the center has already 
increased by three orders of magnitude. The situation is different when
$\Lambda_{-23} = 0.1$. Although the simulation has evolved for 
the same number of cooling times, only those particles which where dense enough to
cool catastrophically ($t_{\rm cool} \la t_{\rm dyn}$) have cooled and
the density has only increased by an order of magnitude. Almost no gas in the
cooling-flow regime ( $t_{\rm dyn} < t_{\rm cool} < t_{\rm Hubble}$) has cooled
down and the total fraction of cooled gas is reduced by a factor of 2.
Increasing the number of dark matter particles without altering the
cooling eliminates this difference (Fig.~\ref{colmod}, right). Again 
all gas within the cooling radius can cool and and settles to a distribution
similar to the $\Lambda_{-23}=0.4$ run.
Since more dynamical
times are now available to react to the loss of pressure support,
and since the central potential cusp of the dark halo is now better
defined, the central gas density increases even more dramatically in
this case. These experiments show
impressively how two--body heating can alter the dynamics and
thermodynamics of the gas, especially in a cooling flow situation.

\section{Summary and Discussion}
We have  analyzed how two--body encounters between fluid elements and dark
matter particles can lead to spurious heating of the gas component
in galaxy formation experiments. We have shown both analytically
and numerically that this process can affect not only the
thermodynamic state, but also the dynamics of the gas. Our analytic
work establishes an upper bound to the mass of a dark matter
particle for two--body heating to be subdominant when simulating
any given physical situation. Our numerical simulations show that
the predicted effect is indeed present, and that the critical mass of 
equation (\ref{masscrit}) provides a reliable guide for designing
numerical experiments so that two--body heating does not cause a
qualitative change in their outcome. Simulations of cooling flows and
of galaxies forming in the presence of a strong UV background are
particularly susceptible to two-body heating, and must therefore be
designed with particular care. 

Although the effect we have discussed is not
important in the catastrophic cooling regime, numerical simulations
have shown that when cooling occurs only in this regime, and no
additional physics is included,
it is not possible to make galaxy disks as large as those observed.
Disk-like objects do form but have too little angular momentum and
so are too concentrated (Navarro \& Benz 1991, Navarro \& White 1994, 
Navarro \& Steinmetz 1996). Furthermore, the observed specific angular 
momenta of giant spiral disks are so large that they must have formed 
late and so most plausibly in the cooling flow regime (\eg White
1991). The solution to this problem may be, as is usually claimed,
that feedback from stellar evolution is an
indispensable ingredient of galaxy formation; such feedback could 
delay the condensation of gas so that most of it cools late,
and with relatively little loss of angular momentum to the dark
matter. Current semi-analytic models include a treatment of such
feedback processes and suggest that a substantial
fraction of the matter in the present galaxy population may have
condensed in the quasi-cooling-flow regime
(White \& Frenk 1991, Kauffmann, Guiderdoni \& White 1993; Fabian
\& Nulsen 1994). Thus, those numerical simulations which, as a result of poor
resolution, avoided excessive catastrophic cooling at early times, may nevertheless have
missed an important ingredient of galaxy formation because of of two-body heating,
in particular, the physics most relevant to the formation of spiral disks.

Our findings lead us to conclude that the common
practice of using the same number of dark matter and gas particles in
cosmological and galaxy formation simulations may be unwise.
The computing time spent per gas particle is usually much higher than for
a dark matter particle, especially if a multiple time-step scheme is
used. It may, therefore, be advantageous to use several times more dark
matter particles than gas particles. The total CPU time will be only moderately
increased while two-body heating can be suppressed by a substantial factor.

Many simulations of individual galaxies and clusters apply a multi-mass
technique (Porter 1985; Katz \& White 1993; Navarro \& White 1994) in which the
tidal field due to surrounding matter is represented by a relatively small
number of massive particles. These massive particles are supposed to stay
outside the high resolution region of interest.  If, however, some of them
accidently pass through a forming galaxy, our experiments demonstrate that they
may prevent cooling of low density gas. It is, therefore, important that
simulations which use such techniques should be checked to ensure that
contamination by massive particles is kept to an acceptably low level.

Although our numerical experiments have used smoothed particle
hydrodynamics, our analytic theory makes no assumption about the 
underlying numerical technique other than that the dark matter is
represented by discrete particles, and that sound waves and shocks
generated in the gas component will be dissipated as heat. As a
result, the considerations of this paper should apply to all the
techniques currently used to simulate cosmological
hydrodynamics and galaxy formation.

\end{document}

%% file: laa.tex
\newcommand{\sun}{\hbox{$\odot$}}
\newcommand{\la}{\mathrel{\mathchoice
{\vcenter{\offinterlineskip\halign{\hfil
$\displaystyle##$\hfil\cr<\cr\sim\cr}}}
{\vcenter{\offinterlineskip\halign{\hfil$\textstyle##$\hfil\cr
<\cr\sim\cr}}}
{\vcenter{\offinterlineskip\halign{\hfil$\scriptstyle##$\hfil\cr
<\cr\sim\cr}}}
{\vcenter{\offinterlineskip\halign{\hfil$\scriptscriptstyle##$\hfil\cr
<\cr\sim\cr}}}}}

\newcommand{\utw}{\smash{\rlap{\lower5pt\hbox{$\sim$}}}}
\newcommand{\udtw}{\smash{\rlap{\lower6pt\hbox{$\approx$}}}}

\newcommand{\diameter}{{\ifmmode\mathchoice
{\ooalign{\hfil\hbox{$\displaystyle/$}\hfil\crcr
{\hbox{$\displaystyle\mathchar"20D$}}}}
{\ooalign{\hfil\hbox{$\textstyle/$}\hfil\crcr
{\hbox{$\textstyle\mathchar"20D$}}}}
{\ooalign{\hfil\hbox{$\scriptstyle/$}\hfil\crcr
{\hbox{$\scriptstyle\mathchar"20D$}}}}
{\ooalign{\hfil\hbox{$\scriptscriptstyle/$}\hfil\crcr
{\hbox{$\scriptscriptstyle\mathchar"20D$}}}}
\else{\ooalign{\hfil/\hfil\crcr\mathhexbox20D}}%
\fi}}
\catcode`\@=11
\long\def\@makecaption#1#2{
 \vskip 10pt
 \setbox\@tempboxa\hbox{#1: #2}
 \ifdim \wd\@tempboxa >\hsize \unhbox\@tempboxa\par \else \hbox
to\hsize{\box\@tempboxa\hfil}
 \fi}

\def\@cite#1#2{{#1\if@tempswa , #2\fi}}
\def\@biblabel#1{\hskip-7truemm}
\catcode`\@=12